\documentclass[aps,pra,showpacs,twocolumn,superscriptaddress]{revtex4}

\usepackage{graphicx}
\usepackage{dcolumn}
\usepackage{bm}
\usepackage {amsmath}

\begin{document}

\title{Entangled state generation with an intrinsically pure single-photon source and a weak coherent source}

\author{Rui-Bo Jin}
\affiliation{Research Institute of Electrical Communication, Tohoku University, Sendai 980-8577, Japan}
\author{Ryosuke Shimizu}
%\affiliation{PRESTO, Japan Science and Technology Agency, Kawaguchi 332-0012, Japan}
\affiliation{Center for Frontier Science and Engineering, University of Electro-Communications, Tokyo 182-8585, Japan }
\author{Fumihiro Kaneda}
\affiliation{Research Institute of Electrical Communication, Tohoku University, Sendai 980-8577, Japan}
\author{Yasuyoshi Mitsumori}
\affiliation{Research Institute of Electrical Communication, Tohoku University, Sendai 980-8577, Japan}
\author{Hideo Kosaka}
\affiliation{Research Institute of Electrical Communication, Tohoku University, Sendai 980-8577, Japan}
\author{Keiichi Edamatsu}
\affiliation{Research Institute of Electrical Communication, Tohoku University, Sendai 980-8577, Japan}

\date{\today }

\begin{abstract}
We report on the experimental  generation of an entangled state with a spectrally pure heralded single-photon state and a weak coherent state.
Our source, which was as efficient as that reported in our previous report [Phys. Rev. A 83, 031805 (2011)],
was much brighter than those reported in earlier experiments using similar configurations.
This entanglement system is useful for quantum information protocols that require indistinguishable photons from independent sources.
\end{abstract}


%03.67.Bg 	Entanglement production and manipulation (for entanglement in Bose-Einstein condensates, see 03.75.Gg)

%\pacs{03.65.Ud, 03.65.Ta, 42.50.Dv, 42.65.Lm}
%03.65.Ud Entanglement and quantum nonlocality
%03.65.Ta Foundations of quantum mechanics; measurement theory
%42.50.Dv Quantum state engineering and measurements
%42.65.Lm Parametric down conversion and production of entangled photons

\maketitle

\section{Introduction}

In many quantum information processing protocols, indistinguishable photons from independent sources are employed as the quantum bits (qubits)\cite{Knill2001, Lanyon2007, Pan2012}.
Quantum operations are performed on them to realize the protocols physically.
Since such quantum operations are often implemented based on quantum interference between photons,
only when the photons are highly indistinguishable can we achieve a high operation fidelity.
However, under current technologies, devising two indistinguishable and independent  single-photon sources is still not an easy task.
To prepare such sources, one can choose two heralded single-photon sources from two spontaneous parametric down-conversion (SPDC) processes \cite{Kaltenbaek2006, Mosley2008a}, or combine one photon from an SPDC source and another photon from a weak coherent source \cite{Rarity2003, Jin2011}.
Generally speaking, the coincidence counts between independent sources of the former scheme is lower than that of the latter scheme.
For example, the coincidence count between independent sources in Refs \cite{Kaltenbaek2006, Mosley2008a} was much smaller than that in Ref \cite{Jin2011}.
In Ref \cite{Jin2011}, we have shown that high visibility can be achieved in the interference by highly indistinguishable photons from a coherent source and an intrinsically pure single-photon source.
In this paper we  demonstrate the application of such indistinguishable photons in  entangled state generation.
We also analyze the generated state taking account of unwanted higher-order photon states;
this analysis essential to entangled photon sources that use a coherent source and a heralded single-photon source.
\section{Method}
\begin{figure}[tbp]
\includegraphics[width= 0.3 \textwidth]{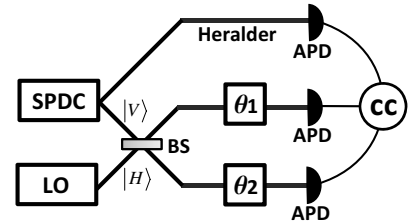}
\caption{(Color online) Schematic model of the experiment. A pure
heralded single photon (V polarization) from SPDC is combined with
a photon (H polarization) from LO at the beam splitter (BS). Then,
their polarization states are analyzed  by two polarizers ($\theta_1$ and $\theta_2$). The second ¡°idler¡± photon from the SPDC is
used as a heralder. Finally, these photons are detected by three
detectors (APD) and recorded by a three-fold coincidence counter
(CC).
 } \label{model}
\end{figure}

The schematic model is shown in Fig.\,\ref{model}.
Here, the heralded, pure single photon from SPDC has vertical (V) polarization, while the LO photon has  horizontal (H) polarization.
Then the two photons are combined at the beam splitter (BS).
After the BS,  the two-photon polarization state is
\cite{Ou1988,Edamatsu2007}
\begin{equation}\label{eq1}
 \begin{array}{l}
 \left| \psi  \right\rangle  = \frac{1}{{\sqrt 2 }}(\left| H \right\rangle _1  + i\left| H \right\rangle _2 ) \otimes \frac{1}{{\sqrt 2 }}(\left| V \right\rangle _2  - i\left| V \right\rangle _1 )\\
 =\frac{1}{2}(\left| H \right\rangle _1 \left| V \right\rangle _2  + \left| V \right\rangle _1 \left| H \right\rangle _2  - i\left| H \right\rangle _1 \left| V \right\rangle _1  + i\left| H \right\rangle _2 \left| V \right\rangle _2 ), \\
 \end{array}
 \end{equation}
where $1$ and $2$ denote the two output modes of the beam splitter.
If we select only the first two terms, by observing the coincidence events only when the two photons split into  the two output ports, the resultant state is
\begin{align}\label{psi+}
\left| \psi ^+  \right\rangle
&= \frac{1}{{\sqrt 2 }}
\left(\left| H \right\rangle _1 \left| V \right\rangle _2  + \left| V\right\rangle _1 \left| H \right\rangle _2 \right)  \nonumber \\
&\equiv \frac{1}{{\sqrt 2 }}
\left(\left| H V \right\rangle   + \left| V H \right\rangle  \right).
\end{align}
This is the maximally entangled Bell state \cite{Note-1}.

\begin{figure}[tbp]
\includegraphics[width= 0.48\textwidth]{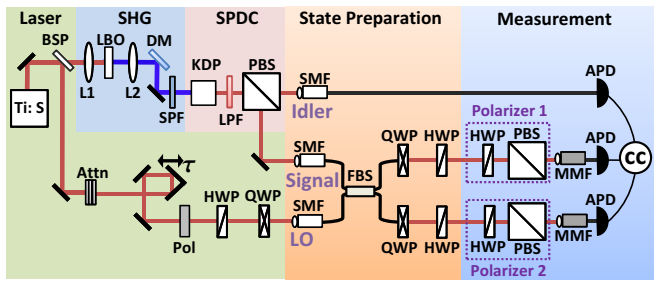}
\caption{(Color online) The experimental setup. BSP=beam sampler,
DM=dichroic mirror, SPF=short wave pass filter, LPF=long wave pass
filter, PBS=polarizing beam splitter, SMF=single-mode fiber,
FBS=fiber beam splitter, QWP=quarter wave plate, HWP=half wave
plate, MMF=multi mode fiber, Pol=polarizer, Attn=attenuator,
APD=avalanche photodiodes, CC=coincidence counter.} \label{setup}
\end{figure}

The experimental setup is shown in Fig.\,\ref{setup}.
Femtosecond laser pulses (temporal duration $\sim$ 150 fs, center wavelength = 830 nm, repetition rate $f= 76$~MHz)
from the mode-locked Titanium sapphire laser (Coherent, Mira900) were frequency-doubled by an 0.8-mm-thick lithium triborate (LBO) crystal and were used as the pump source for the SPDC.
Pump pulses with power of 60 mW passed through a 15-mm-long KDP crystal cut for type-II (eoe) degenerate phase-matching for the SPDC at 830 nm.
The down-converted photons, i.e., the signal (o-ray) and idler (e-ray) were separated by a polarizing beam splitter (PBS).
Then, idler photons were coupled into a single-mode fiber, and signal photons were coupled into a 50:50 single-mode fiber beam splitter (FBS) (Thorlabs, FC830-50B-FC).
Note that the polarization of the signal photon was $|V\rangle$  in this configuration.
Fundamental laser pulses reflected from a beam sampler and highly attenuated by neutral density filters were used as the LO photons.
The polarization of the LO photon was adjusted to be $|H\rangle$   or   $|V\rangle$   by a polarizer, a half-wave plate (HWP) and a quarter-wave plate (QWP),
so that the polarization of LO was either in parallel or orthogonal  to that of the signal photons at the FBS.
Since the polarization mode changes in the fiber, the output photons from FBS were compensated by two sets of QWP and HWP; then, they were tested by two polarizers, consisting of an HWP and a PBS.
Finally, all the collected photons were sent to three silicon avalanche photodiode (APD) detectors (PerkinElmer, SPCM-AQRH14) connected to a three-fold coincidence counter.
In our typical experimental condition, the observed single count rates ($C_1$) of signal and idler photons were both
10~kcps, and the coincidence count rate ($C_2$) between the signal and idler was 3~kcps.
The count rate  ($C_l$) of  LO photons was 600~kcps, and the three-fold coincidence count rate 
($C_3$) after the FBS 
was 6~cps.
Note that $C_3$ is expected to be 
$C_3 \sim C_2 C_l /(2f) \sim 11$~cps,
%where $f$ is the pump repetition rate, 
showing reasonable agreement with the observed value.
Taking account of 
the total collection efficiencies ($\eta=0.3$ for signal and idler photons and $\eta_l=0.37$ for LO photons), 
which includes the fiber-coupling efficiency
% (0.5 for signal and idler photons, 0.75 for LO photons) 
and the detection efficiency,
% (0.5), 
the mean photon numbers per pulse were estimated to be 
 $ \mu = C_1/(\eta f) \sim 4.3 \times 10^{ - 4} $ for signal and idler photons, and 
 $\nu = C_l /(\eta_l f) \sim 2.1 \times 10^{ - 2} $ for LO photons.

\section{Results}
\begin{figure}[tbp]
\begin{center}
\includegraphics[width=0.3\textwidth]{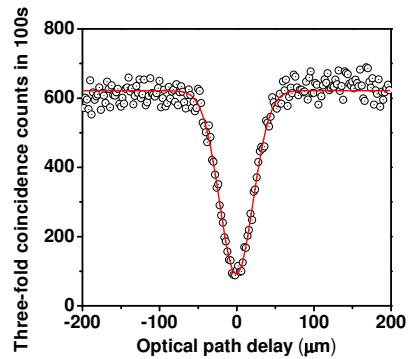}
\caption{(Color online) Experimental result of three-fold coincidence
counts as a function of the optical path delay, with a visibility
of $85.4 \pm 0.3\% $ with no background subtraction.}
\label{3folddip}
\end{center}
\end{figure}

We first carried out three-fold Hong-Ou-Mandel (HOM) interference \cite{Hong1987} to test the indistinguishability between the LO and the heralded signal.
We set both LO and signal photons in $|V\rangle$ and recorded the three-fold coincidence counts as a function of the optical path delay $\tau$ between the signal and the LO.
Figure\,\ref{3folddip} shows the result of the three-fold HOM interference, which exhibits a clear HOM dip at $\tau=0$  with a visibility of 85.4 $\pm$ 0.3\% and an FWHM of 50.7 $\mu$m.
From the result, we confirm  the high indistinguishability between LO and the heralded signal.
\begin{figure}[tbp]
\includegraphics[width=0.48 \textwidth]{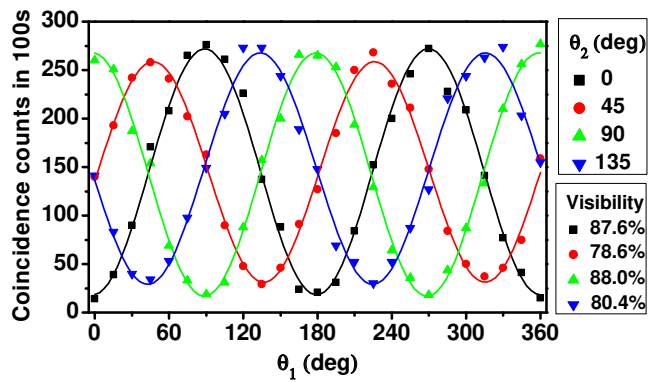}
\caption{(Color online) Experimental result of three-fold coincidence
counts as a function of Polarizer 1 ($\theta_1$) and Polarizer 2
($\theta_2$). The data have no background subtraction.}
\label{interference}
\end{figure}

Next, we demonstrate the violation of the Bell inequality using this source.
We set the signal and LO polarizations as $|V\rangle$ and $|H\rangle$, respectively, and adjusted the optical path delay to $\tau=0$.
We carried out a polarization correlation measurement by recording the coincidence counts while changing the angles $\theta_1$ and $\theta_2$  of Polarizer 1 and Polarizer 2, respectively.
The experimental results for some fixed values of $\theta_2$ ( $\theta_2 = 0$, 45, 90 and $135^\circ$) are shown in Fig.\,\ref{interference}.
As indicated in the figure, we observed that the fringe visibilities were higher than 78.6\%, which exceeded 71\%, the bound required to  violate the Bell-CHSH inequality \cite{Clauser1978}.
We obtained the Bell parameter $S$, which directly indicated the violation of the Bell  inequality \cite{Clauser1969}.
In this experiment, $S$ is given as
\begin{equation}\label{s}
S = |E(\theta_1,\theta_2) + E(\theta'_1,\theta_2)+
     E(\theta_1,\theta'_2)
     - E(\theta'_1,\theta'_2)|,
\end{equation}
where $E(\theta_1,\theta_2)$ is given by
\begin{align}\label{e}
&E(\theta_1,\theta_2)   \nonumber\\
&= \frac{C(\theta_1,\theta_2) +
C(\theta_1^{\bot},\theta_2^{\bot}) - C(\theta_1^{\bot},\theta_2)-
C(\theta_1,\theta_2^{\bot})}{C(\theta_1,\theta_2) +
C(\theta_1^{\bot},\theta_2^{\bot}) + C(\theta_1^{\bot},\theta_2)+
C(\theta_1,\theta_2^{\bot})},
\end{align}
where $C(\theta_1,\theta_2)$ is the coincidence count for different polarization angles and $\theta_i^{\bot}\equiv\theta_i+90^\circ$.
The observed data to obtain $S$ is shown in Table \ref{table1}.
The obtained value of $S$ was 2.23 $\pm$ 0.03 $>$ 2, which demonstrated a clear violation of Bell inequality by more than 7 times the standard deviation.
The slight degradation of the $S$ value (ideally $S=2\sqrt2$) might have originated from incomplete compensation of the possible polarization rotation in the fiber.
\begin{table}[t]
\begin{ruledtabular}
\begin{tabular}{c|cccc}
 $ \theta_2 \backslash\, \theta_1 $ &$45^\circ$ & $90^\circ$ & $135^\circ$ & $180^\circ$\\
 \hline\\
  $22.5^\circ$  & 945  &  960   &  261     & 276\\
  $67.5^\circ$  & 910  &  234   &  319     & 998\\
  $112.5^\circ$ & 310  &  238   &  952     & 1020\\
  $157.5^\circ$ & 269  &  989   &  901     & 263 \\
\end{tabular}
\end{ruledtabular}
\caption{\label{table1}Coincidences counts $C(\theta_1,\theta_2)$
as a function of different angles of polarizers.}
\end{table}

\begin{figure}[tbp]
\includegraphics[width= 0.48\textwidth]{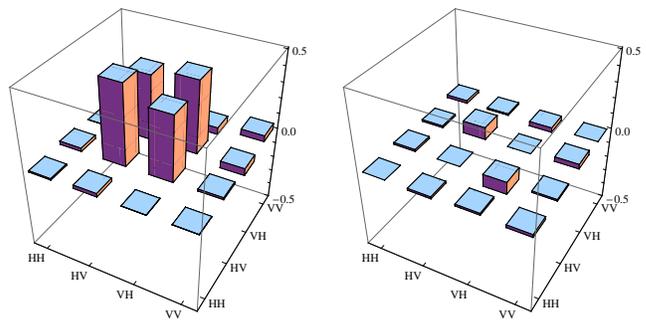}
\caption{(Color online) Real (left) and imaginary (right) parts of
the reconstructed density matrix.} \label{matrix}
\end{figure}

We also carried out state tomography \cite{James2001} of our two-photon polarization state.
Polarizers 1 and  2  in Fig.\,\ref{setup} were replaced by the combinations of HWP, QWP and PBS, to allow polarization correlation analysis in not only linear but also circular polarization bases.
The density matrix $\rho_{exp}$ reconstructed with a maximum likelihood estimation method \cite{James2001} is shown in  Fig.\,\ref{matrix}.
We see that the reconstructed density matrix is close to the ideal one expected from Eq. (\ref{psi+})
\begin{equation}\label{1}
\begin{array}{cc}
\left| \psi ^+  \right\rangle \left\langle \psi ^+  \right|=  &  \frac{1}{2} (\left| HV \right\rangle \left\langle HV \right|  + \left| VH \right\rangle \left\langle VH \right|    \\
     & + \left| HV \right\rangle \left\langle VH \right|   +   \left| VH \right\rangle \left\langle HV \right|   ).  \\
\end{array}
\end{equation}
The small imbalance between
$\left| HV \right\rangle \left\langle HV \right|$ and $\left| VH \right\rangle \left\langle VH \right|$
was mainly attributable to the imbalance between the reflectance and transmittance of the FBS used.
Small components in the imaginary part of the matrix elements
$\left| HV \right\rangle \left\langle VH \right|$ and $\left| VH \right\rangle \left\langle HV \right|$
indicate that the generated state underwent a small phase change between
$\left| HV \right\rangle$ and $\left| VH \right\rangle$
arising from imperfect cancellation of the fiber birefringence.
Also, partial distinguishability between the signal and LO photons,
which was already observed as the imperfect visibility in Fig.\,\ref{3folddip},
 degraded the coherence between
$\left| HV \right\rangle$ and $\left| VH \right\rangle$ and resulted in the slightly smaller values of the
$\left| HV \right\rangle \left\langle VH \right|$ and $\left| VH \right\rangle \left\langle HV \right|$
components.
Nevertheless,
the calculated value of fidelity \cite{Jozsa1994}, $F \equiv \left\langle {\psi ^ +} \right|\rho _{\exp } \left| {\psi ^ +  } \right\rangle $, to the ideal Bell state $|\psi^+\rangle$ was estimated as 
0.88  $\pm$ 0.01.
%88.3 $\pm$ 0.9 \%.
%
The corresponding concurrence and entanglement of formation (EOF) \cite{Wootters1998} are $0.793$ and $0.712$, respectively.
These values indicate that our obtained state was highly entangled.

\section{Discussion}

Thus far, we have implicitly assumed that three single photons, one in the signal mode, one in the idler mode, and one in the LO mode, contribute to our measurement.
This assumption is valid only when the mean number of photons in each mode is sufficiently low.
However, in practice,  we should consider the effects of higher-order photon states \cite{Wieczorek2009, Laskowski2009}.
The state of the signal and idler photons emitted from the SPDC can be expressed on the basis of photon number as
\begin{equation}\label{a}
\left| \varphi  \right\rangle
= a_0\left| {00} \right\rangle
+ a_1 \left| {11} \right\rangle
+ a_2 \left| {22}\right\rangle
+ ...\, ,
\end{equation}
where $|nn\rangle\equiv|n\rangle_s\otimes|n\rangle_i$ denotes the $n$-pair state containing $n$ photons in both signal and idler modes.
The photon number probability follows the geometric distribution given by $ \left| a_n \right|^2=p(1-p)^n$, where $p=1/\left( 1+ \mu \right) $ and $\mu$ is the mean photon number in the signal (or idler) mode.
For  $ \mu = 4.3 \times 10^{ - 4} $,
$|a_0| ^2 $, $|a_1| ^2$  and $|a_2| ^2$ are estimated as 0.99957, $ 4.3 \times 10^{ - 4} $ and   $ 1.8 \times 10^{ - 7} $, respectively.
The two-pair component is 3 orders of magnitude lower than the single-pair component.
The weak coherent state of LO can be written as
\begin{equation}\label{c}
\left| \alpha  \right\rangle
= c_0 \left| 0 \right\rangle
+ c_1 \left| 1 \right\rangle
+ c_2 \left| 2 \right\rangle
+ ...\, ,
\end{equation}
where  $|n\rangle$ represents the $n$-photon state in the LO mode.
The photon number probability follows the Poisson distribution given by $\left| c_n  \right|^2 = e^{-\nu} \nu^n / n!$, where $\nu$ is the mean photon number in the LO mode.
For $\nu=2.1 \times 10^{-2}$,
$|c_0| ^2$, $|c_1| ^2$ and $|c_2| ^2$ were estimated as 0.979, $ 2.1\times10^{-2}$ and $ 2.2\times10^{-4}$, respectively.
The two-photon component is 2 orders of magnitude lower than the single-photon component.
Therefore, the higher-order components are much smaller than the single-pair or the single-photon component in both SPDC and LO, and the interference between LO and the heralded signal can safely be regarded as the interference between the two single-photon states.

More specifically, the three-fold coincidence events of our signal can be mainly contributed to three combinations of the states:
$\left|{11} \right\rangle  \otimes \left| 1 \right\rangle $,
$\left| {11} \right\rangle  \otimes \left| 2\right\rangle $
and
$\left| {22} \right\rangle   \otimes \left| 0 \right\rangle$,
where $\left|{nn} \right\rangle$ and  $\left| n \right\rangle$ are the states in (\ref{a}) and (\ref{c}), respectively.
Assuming that the total collection efficiencies are $\eta$ for  the signal and idler and $\eta_l$ for LO,
the probabilities of the three-fold coincidences occurring at these three states are
\begin{align}
P_{111}&= \frac12 |a_1|^2 |c_1|^2 \eta^2\eta_l, \label{p1}\\
P_{112}&= \frac12 |a_1|^2 |c_2|^2 \eta \eta_l^2, \label{p2}\\
P_{220}&= \frac12 |a_2|^2 |c_0|^2\left(1-\bar\eta^2\right) \eta^2, \label{p3}
\end{align}
respectively, where $\bar\eta=1-\eta$.
Note that the factors for the polarization measurement are omitted from (\ref{p1})-(\ref{p3}).
Only when $P_{111}$ is much larger than $P_{112}$ and $P_{220}$ can the $\left| {11}\right\rangle  \otimes  \left| {1}\right\rangle$ state dominates the coincidence events, and thus our signal can be what we expect.
In our case, we estimate that $\left(P_{112}+P_{220}\right)/P_{111} = 0.040$, which confirms that the above condition is fulfilled.
The estimated fidelity $F$ assuming that these higher-order terms are added to the ideal Bell state $|\psi^+\rangle$ is 0.96, 
which is much higher than that we observed (0.88).
Thus, we conclude that the higher-order terms are not the major origins for the degradation of fidelity (and entanglement) of our state.
The degradation is most likely caused by partial distinguishability of the single photon state we generated, as indicated by its incomplete HOM interference (Fig.~\ref{3folddip}).

It is worth discussing the efficiency of our source in comparison with that of the previous report that also used independent photon sources to generate entangled photons \cite{Pittman2003a}.
In our experiment, we obtained the three-fold coincidence count rate of $\sim$6 cps,
which is approximately 60 times larger than that previously reported \cite{Pittman2003a}.
Since our source eliminated the use of narrow bandpass filters to make the signal and LO photons indistinguishable, our source is in principle much more efficient than those using bandpass  filters.
This is the intrinsic advantage of our heralded single-photon source that uses group-velocity matching to generate spectrally pure single photons.

It is also worth comparing the efficiency of our system with that of other systems that make use of a pair of heralded single-photon sources generated by SPDC.
In general, it is a difficult task to increase the mean photon number generated by SPDC up to the order of 0.1.
Thus, systems using a pair of SPDCs have an intrinsic drawback in their generation efficiency, unless they use wave-guided SPDC \cite{Zhong2012} or high-power lasers to increase the SPDC efficiency.
In addition, this scheme requires four-fold coincidence events and thus the total collection efficiency is proportional to $\eta^4$, which again decreases the observed event rate.
For instance, in Refs \cite{Kaltenbaek2006, Mosley2008a}, the four-fold coincidence rate was less than 0.3 cps.
In our system, on the other hand, one can easily optimize the mean photon number of LO, within the range where the above-mentioned condition is satisfied.
We eventually obtained a much higher  three-fold event rate (6 cps)  than the previous four-fold ones.
Also, the experimental setup of our system is obviously much easier, since we only need one SPDC crystal.
Thus, our system has an advantage in efficient generation of not only entangled photons but also indistinguishable photons from independent sources.

\section{Conclusion}

In summary, we have demonstrated the generation of a polarization-entangled state and violation of Bell  inequality with a spectrally pure,
heralded single-photon source and a weak coherent source.
Our system for producing entangled photons is much brighter than those reported in earlier experiments that made use of a pair of heralded single-photon sources.
We have characterized the generated state using the Bell inequality test and the state tomography.
We have also analyzed the state taking account of unwanted higher-order photon states.
These analyses indicate that the obtained state was highly entangled with negligible contribution from the higher-order state.
This entanglement system will be useful for  quantum information protocols which require indistinguishable photons from independent sources.

\section{Acknowledgements}

This work is supported by a Grant-in-Aid for Creative Scientific Research (17GS1204) from the Japan Society for the Promotion of Science.

%\bibliography{bellinequjin}

%\begin{thebibliography}{99}
%
%\bibitem{Note-1}
%{\it In some literature such as Ref.~\cite{Pittman2003a},
%the resultant state is reported to be $ \left| \psi^-  \right\rangle  = \frac{1}{{\sqrt 2 }}\left(  \left| {HV} \right\rangle  - \left| {VH} \right\rangle \right)$,
%which is different from Eq.~(\ref{psi+}).
%%
%This difference is attributed to different definitions in the electric field directions for the reflected photons after the beamsplitter \cite{Edamatsu2007}. }
%
%
%
%\end{thebibliography}

\end{document}